# New type of incommensurate magnetic ordering in Mn$_3$TeO$_6$

S.A. Ivanov[ab], P. Nordblad[b], R. Mathieu[b*], R. Tellgren[c],
C. Ritter[d], N.V. Golubko[a], E.D. Politova[a], M. Weil[e]

a- Department of Inorganic Materials, Karpov' Institute of Physical Chemistry, Vorontsovo pole, 10 105064, Moscow K-64, Russia
b- Department of Engineering Sciences, Uppsala University, Box 534, SE-751 21 Uppsala, Sweden
c- Department of Materials Chemistry, Uppsala University, Box 538, SE-751 21 Uppsala, Sweden
d- Institute Laue Langevin, Grenoble, France
e- Institute for Chemical Technologies and Analytics, Vienna University of Technology, A-1060 Vienna, Austria

## Abstract

The complex metal oxide Mn$_3$TeO$_6$ exhibits a corundum related structure and has been prepared both in forms of single crystals by chemical transport reactions and of polycrystalline powders by a solid state reaction route. The crystal structure and magnetic properties have been investigated using a combination of X-ray and neutron powder diffraction, electron microscopy, calorimetric and magnetic measurements. At room temperature this compound adopts a trigonal structure, space group $R\bar{3}$ with $a$ = 8.8679(1) Å, $c$ = 10.6727(2) Å. A long-range magnetically ordered state is identified below 23 K. An unexpected feature of this magnetic structure is several types of Mn-chains. Under the action of the incommensurate magnetic propagation vector k = [0, 0, 0.4302(1)] the unique Mn site is split into two magnetically different orbits. One orbit forms a perfect helix with the spiral axis along the c-axis while the other orbit has a sine wave character along the c-axis.

## 1. Introduction

Magnetoelectric multiferroics are materials that not only show ferroelectric and magnetic order simultaneously, but also display coupling between these properties [1, 2]. Such materials have been studied since more than four decades [3-5]. Initially only materials with weak magnetoelectric coupling were found and the activity in this research field remained marginal for many years [6]. However, a boosted interest arose after the discovery of multiferroic materials showing a large magnetoelectric effect [7, 8]. A key factor in the ferroelectricity of these materials lies in their noncollinear spiral magnetic structures with a cycloidal component in which the magnetic structure itself breaks inversion symmetry [9, 10]. So far, only a few single-phase multiferroics have been discovered [11]. A class of potential single-phase multiferroic materials possesses geometrically frustrated spin networks, which prevent the formation of conventional collinear spin structures. In this class of antiferromagnetic materials, an incommensurate magnetic structure with spiral spin order can result, and the magnetic phase transition to this non-centrosymmetrically ordered magnetic state can induce a spontaneous electric polarization via the spin-orbit coupling [12].

\* Corresponding author.
*E-mail address:* roland.mathieu@angstrom.uu.se

To date, much of the search for new multiferroic materials has been focused on perovskite-based materials [13-15]. However, the coexistence of ferromagnetism and ferroelectricity in single-phase perovskites is hampered by the conflicting electronic requirements of the two ground states - $d^n$ ions are required for ferromagnetism, but $d^0$ ions support the out of center distortions seen in classical perovskite ferroelectrics [2]. The spatial separation of ions supporting the magnetization and the polarization is an attractive method to overcome this problem, but it requires a complex crystal structure with distinct sites for the magnetically and electrically active ions.

Several complex metal oxides $A_3TeO_6$ (ATO, A=Mn, Co, Ni and Cu) have attracted renewed interest in the research community due to their low temperature magnetic properties [16-18]. The structure types of the different ATOs including $Mn_3TeO_6$ (MTO) are similar to the perovskite structure; however, the presence of new crystallographically non-equivalent sites for magnetic cations provides extra degrees of freedom to manipulate the structure. $Mn_3TeO_6$ was first described and prepared in microcrystalline form by Bayer [19] who found the compound to be isotypic with the corundum related $Mg_3TeO_6$ family. Later, Kosse et al. [20,21] identified several $A_3TeO_6$ oxides as ferroelectrics. Here we report results from the first neutron diffraction experiments on $Mn_3TeO_6$. The NPD experiments have been made on powder samples at different temperatures and it is found that the compound exhibits long-range incommensurate magnetic order at low temperatures.

## 2. Experimental

Single crystals and ceramic samples of $Mn_3TeO_6$ were synthesized. A mixture of MnO and $TeO_3$ in the stoichiometric ratio 3:1 was thoroughly ground, pressed into a pellet and placed in a silica ampoule which was evacuated, sealed, and heated within 3 h to 1103 K and kept at this temperature for 3 days. X-ray powder diffraction of the light-brown microcrystalline material revealed a single phase product. 80 mg of this material was mixed with 5 mg $PtCl_2$ and loaded in an evacuated and sealed silica ampoule which was heated in a temperature gradient from 1103 to 1023 K. At this temperature, $PtCl_2$ decomposes with release of $Cl_2$ which then serves as the transport agent. After 5 days, amber coloured crystals of the title compound with a plate-like habit and an edge-length up to 0.8 mm had formed in the colder part of the ampoule. A high quality ceramic sample of MTO was prepared by a conventional solid-state ceramic route. High purity $Mn_2O_3$ and telluric acid $H_2TeO_4 \cdot 2H_2O$ were used as starting materials. The raw materials were weighed in appropriate proportions for the $Mn_3TeO_6$ composition. The homogenized stoichiometric mixtures were calcined at 570 K for 7 h, ground into fine powders, pressed and annealed again several times with a temperature interval of 100 K up to 1270 K with intermediate milling.

The chemical compositions of the prepared crystals and ceramic samples were determined by energy-dispersive spectroscopy (EDS) using a JEOL 840A scanning electron microscope and INCA 4.07 (Oxford Instruments) software. The analyses performed on several samples showed that the concentration ratios of Mn:Te were as expected for $Mn_3TeO_6$ within the instrumental resolution (0.05). Structure analysis of MTO single crystals [22] was performed at room temperature on a SMART Bruker three-circle diffractometer.

The phase identification and purity of the powder samples were checked from X-ray powder diffraction patterns (XRPD) obtained with a Bruker D-5000 diffractometer using Cu $K_\alpha$ radiation. For that purpose, the ceramic MTO samples were crushed into powder in an agate mortar and suspended in ethanol. A Si substrate was covered with several drops of the resulting suspension, leaving randomly oriented crystallites after drying. The XRPD data for Rietveld analysis were collected at room temperature on a Bruker D8 Advance diffractometer (Vantec position-sensitive detector, Ge-monochromatized Cu K$_\alpha$ radiation, Bragg-Brentano geometry, DIFFRACT plus software) in the 2$\theta$ range 10-152° with a step size of 0.02° and a counting time of 15 s per step. The slit system was selected to ensure that the X-ray beam was completely within the sample for all 2$\theta$ angles. Polycrystalline MTO material was characterized by SHG measurements in reflection geometry, using a pulsed Nd:YAG laser ($\lambda$=1.064 μm). The SHG signal $I_{2\omega}$ from the sample was measured relative to an α-quartz standard at room temperature in the Q-switching mode with a repetition rate of 4 Hz.

The magnetization experiments were performed in a Quantum Design MPMSXL 5 T SQUID magnetometer. The magnetization (M) of single-crystal and ceramic samples was recorded as a function of temperature T in the interval 5-300 K in 20 and 1000 Oe field using zero-field-cooled (ZFC) and field-cooled (FC) protocols. The magnetization of single-crystal material was also recorded as a function of the magnetic field, up to 5 Tesla at low temperatures (2K). Specific heat measurements between 2 K and 100 K were performed using a relaxation method on a Physical Properties Measurement System (PPMS6000) from Quantum Design Inc.

The neutron scattering lengths of Mn and Te are very different, thus the chemical composition can be observed by neutron powder diffraction (NPD) with good precision. The neutron scattering length of oxygen is comparable to those of the cations and NPD provide accurate information on its position and stoichiometry. The neutron diffraction experiments on the MTO samples were performed at the Institute Laue-Langevin (Grenoble, France) on the powder diffractometer D1A ($\lambda$ = 1.9095(1) Å) in the 2$\theta$-range 10– 156,9° with a step size of 0.1°. The powdered sample was inserted in a cylindrical vanadium container. A helium cryostat was used to collect data in the temperature range 1.5-295 K. Nuclear and magnetic refinements were performed by the Rietveld method using the FULLPROF software [23]. The diffraction peaks were described by a pseudo-Voigt profile function, with a Lorentzian contribution to the Gaussian peak shape. A peak asymmetry correction was made for angles below 35° (2$\theta$). Background intensities were estimated by interpolating between up to 40 selected points (low temperature NPD experimental data) or described by a polynomial with six coefficients. Analysis of the coordination polyhedra of the cations was performed using the IVTON software [24]. The magnetic propagation vector was determined from the peak positions of the magnetic diffraction lines using the K-search software which is included in the FULLPROF Suite package [23]. Magnetic symmetry analysis was then made using the program BASIREPS [25] which calculates the allowed irreducible representations and their basis vectors. The proposed models were one by one tested against the measured data. The variant for which the structural refinement was stable and the reliability factors at a minimum was chosen as the final model.

## 3. Results

### 3.1 Structural Characterization

It was found that phase formation of $Mn_3TeO_6$ starts in the temperature range between 870 and 920 K, and single-phase samples have been obtained after annealing at 970 K. According to the elemental analyses performed on 20 different crystallites, the metal compositions of MTO is $Mn_{2.98(2)}Te_{1.02(2)}$, if the sum of the cations is assumed to be 4. These values are very close to the expected ratios and permit to conclude that the sample composition is the nominal one. The oxygen content, as determined by thermogravimetric analysis, is also in agreement with the $Mn_3TeO_6$ composition.

The microstructure of the obtained powders, observed by scanning electron microscopy, reveals uniform and fine grain distribution. The first crystallographic characterization of MTO compound was performed by XRPD analysis at room temperature which showed that the prepared samples are single phase and show very narrow diffraction peaks without any splitting or extra reflections. The result is consistent with that reported for the structure refinement of $Mn_3TeO_6$ based on single crystal data [22]. All patterns could be successfully refined by the Rietveld method. The Mn-O and Te-O bond lengths calculated from the refined lattice parameters and atomic coordinates are in good agreement with earlier observed. Furthermore, the corresponding bond valence sum calculations are consistent with the presence of $M^{2+}$, $Te^{6+}$ and $O^{2-}$ ions.

Second harmonic generation (SHG) measurements at room temperature gave a negative result, thus testifying that at this temperature the MTO compound probably possesses a centrosymmetric crystal structure. This sample still could be non-centrosymmetric, but at a level detectable only with sensitivities beyond $10^{-2}$ of quartz [26].

### 3.2 Magnetic properties

Figure 1(a) shows the temperature dependence of the magnetic susceptibility recorded in magnetic fields of 20 Oe for the single crystal and the ceramic sample used in NPD experiments. The maximum of M/H(T) at 25 K suggests antiferromagnetic ordering with a transition temperature of about 22 K (estimated from the maximum slope of M/H x T vs T) for both the single crystal and ceramic sample. At low temperatures, a relatively sharp peak is observed at 23.5 K in the heat capacity curve (plotted as C/T), as seen in Fig. 1(b). This peak reflects the antiferromagnetic transition detected in the magnetic susceptibility measurements. This latter determination of the Néel temperature is more accurate that the magnetic measurements, as closer data points were recorded in the heat capacity measurements. The temperature dependence of the magnetic susceptibility was also recorded in larger magnetic field, namely 1000 Oe. As seen in Fig. 1 (c), the high-temperature data (50 < T < 300 K) obeys a Curie-Weiss law M/H(T)=C/(T+θ) with a value of θ ~ -120 K and an effective bohr magneton number p of 5.93 $\mu_B$, in very good agreement with the theoretically expected value for $Mn^{2+}$ in a $3d^5$ configuration (S=5/2, L=0, p=5.92 $\mu_B$). The large |θ| value with respect to the antiferromagnetic transition (yielding a frustration parameter f=-θ/$T_N$ ~ 5) indicates a large magnetic frustration. Figure 1 (d) shows the typical magnetic field dependence of the magnetization of the system at low temperatures, also suggesting an antiferromagnetic configuration.

### 3.3 Neutron powder diffraction at room temperature

We started to refine the crystal structure of MTO sample using NPD data at 295 K. To test the space group found from the refinements of the XRD data several centrosymmetric trigonal space groups were initially considered. Rietveld refinements were carried out in all space groups, but a clearly superior fit was obtained using space group $R\bar{3}$ including the structural model refined previously from single crystal XRD data. The obtained atom positions are very similar, but we were able to determine more accurately the oxygen positions due to the characteristics of the neutron scattering. No vacancies were observed in the cationic or in the anionic substructures. Accordingly, the chemical composition seems to be very close to the nominal one and therefore, the Mn oxidation state can be assumed to be +2. The atomic coordinates and other relevant parameters are gathered in Table 1. Selected bond lengths and angles are listed in Table 2. The NPD pattern at 295 K is depicted in Fig. 2. Polyhedral analysis of the $Mn^{2+}$ and $Te^{6+}$ cations in $Mn_3TeO_6$ at different temperatures is presented in Table 3. The crystal structure of MTO is isotypic with $Mg_3TeO_6$ and can be derived from a close packing of strongly distorted hexagonal oxygen layers parallel to (001), with Mn and two distinct Te atoms in the octahedral interstices (see Fig. 3). Both $TeO_6$ octahedra exhibit 3-fold symmetry and are fairly regular (see Table 2), with an average Te-O distance of 1.927 Å, which is in good agreement with the average Te-O distances of other oxotellurates [27]. Each $TeO_6$ octahedron shares edges with six $MnO_6$ octahedra but none with other $TeO_6$ octahedra. Each $MnO_6$ octahedron shares four edges with adjacent $MnO_6$ octahedra, one edge with a $Te(1)O_6$ and another edge with the $Te(2)O_6$ octahedron. The shared edges of the $TeO_6$ octahedra have somewhat shorter O-O distances than the non-shared edges. This is may be due to the high valence of tellurium which has the tendency to keep as far as possible from the $Mn^{2+}$ cations. Each of the two crystallographically independent O atoms is coordinated by one Te and three Mn atoms in a distorted tetrahedral manner. The $MnO_6$ octahedron is considerably distorted which is reflected by the variation of the Mn-O distances between 2.111(6) to 2.374(6) Å (Table 2).

**3.4 Low-Temperature Neutron Powder Diffraction**
Additional NPD measurements were performed at 60 K in order to check whether any crystallographic phase transition occurred in MTO, and at 5 K in order to search for magnetic diffraction peaks. The diffraction profiles are shown in Fig. 4. As discussed in relation to the magnetic properties, the magnetic susceptibility measurements indicate an initial antiferromagnetic transition at about 23 K. Above $T_N \approx 23$ K, the patterns show only the peaks expected from the room-temperature crystal structure. In addition, a diffuse scattering is present on the pattern registered at 60K over a notable $2\theta$ range centered at $2\theta$ around $20°$. This diffuse scattering disappears on cooling below 23 K. Therefore, the diffuse scattering may indicate the presence of short-range antiferromagnetic correlations that develop above $T_N$. The NPD pattern of MTO obtained below the Neel temperature shows the presence of additional peaks of magnetic origin that has to be taken into account for the refinement of the high resolution data at T = 5 K. The magnetic structure and therefore the magnetic symmetry are, however, unknown and have to be determined either by trial and error or by magnetic symmetry analysis [23, 25]. As the magnetic Bragg peaks can not be indexed using simple multiples of the crystallographic unit cell the magnetic structure is incommensurate with respect to the nuclear structure of the crystal. Possible magnetic propagation vectors were determined from the peak positions of the magnetic satellite reflections using the program "K-search" which is part of the FullProf_Suite [23].
A relatively simple value of the propagation vector, which was able to index all the magnetic satellite peaks was obtained as $k = [0, 0, 0.4302(1)]$. The incommensurate component of the propagation vector suggests that the spin interactions are frustrated along the c-direction. It is well known that sufficiently strong frustration in a magnet results in a large number of quasi-degenerate low-energy states that can compete for the ground state.
While the magnetic propagation vector determines the modulation going from one unit cell to another, magnetic symmetry analysis is needed to determine the coupling between the symmetry related magnetic sites within one crystallographic unit cell. The program "BasIreps" was used to find allowed symmetry couplings in the form of irreducible representations and their respective basis vectors. It turned out that the general position in the crystallographic space group R-3 with multiplicity 18, where the Mn site is found, splits under the action of the incommensurate magnetic propagation vector into two independent orbits each having 9 symmetry related sites. Sites, which are crystallographically linked through the inversion centre, are no longer coupled from the magnetic point of view.

For both of these orbits there are three possible irreducible representations each having three basis vectors:

**IReps1** : $Mn_1=(u,v,w)$; $Mn_2=(v,-u-v,w)$; $Mn_3=(-u-v,u,w)$

**IReps2** : $Mn_1=(u,v,w)$; $Mn_2=(R_0+iR_1)\cdot(-v,u+v,-w)$; $Mn_3=(R_0-iR_1)\cdot(u+v,-u,-w)$

**IReps3** : $Mn_1=(u,v,w)$; $Mn_2=(R_0-iR_1)\cdot(-v,u+v,-w)$; $Mn_3=(R_0+iR_1)\cdot(u+v,-u,-w)$

with $R_0 = 0.5$ and $R_1 = 0.866$.

In the above used formalism u, v and w are pointing along the crystallographic a, b and c axis directions. The R-centering couples each of the symmetry related $Mn_{(1,2,3)}$ positions to two further positions which have a spin orientation and a moment value directly determined by the relation between the translations induced by the centering and the propagation vector.
As mentioned above, the two orbits of 9 Mn sites each are independent and can therefore follow different irreducible representations. This leads for the refinement of the magnetic structure to 2 (number of orbits) times 3 (number of basis vectors) independent coefficients. The relative phasing between the modulations of the two orbits has to be determined through the refinement of a phase factor.
All possible combinations of the three allowed irreducible representations for the 2 orbits were tested against the measured data. The refinement program FullProf allows to refine either purely real coefficients which leads to a purely sinusoidal modulation of the moment values or a refinement taking into account the imaginary components of the basis vectors which leads to a helical type arrangement of the spins [28].
Rietveld refinements on the powder diffraction data, collected at 5 K, quickly showed that only with the 2-dimentional representation IReps2 (with real and imaginary components), a succesfull refinement of the data can be obtained. A spiral modulation of spins within the hexagonal basal plane gives significantly lower agreement indices compared to a purely sinusoidal modulation. Models assuming a purely sinusoidal modulation lead furthermore to unrealistic high values of the magnetic moments and were discarded. Table 4 shows the refined coefficients for the two Mn orbits. It can be seen that the coefficients $C_1$ and $C_2$ describing the magnetic moments within the basal hexagonal plane can be forced to have the same value without affecting the goodness of the refinement. The final fit of the Rietveld refinement is shown in Fig. 5. With either $C_1$ or $C_2$ being imaginary this would lead in an orthogonal system to a perfect helix structure. Due to the hexagonal axes with an angle of 120° the resulting spin structure has in the case of MTO an elliptical envelope perpendicular to the propagation vector. This is illustrated in Fig. 6 (a). The helical spin evolution of Mn(1) along the c-axis can be seen in Fig. 6 (b). While the first orbit of the Mn sites, Mn(1) has no spin component in the c-direction as $C_3=0$, the second orbit Mn(2) has its main spin component in the c-direction with the components within the basal plane strongly reduced. This can be seen in Fig. 6 (c). The sine wave character of the Mn(2) moments is shown in Fig. 6 (d), where only the component along the c-axis is plotted. We recall that for the final refinement of the magnetic structure only 3 coefficients plus the phase factor are needed.
Magnetic moment values are in the range 3.6(2) $\mu_B$ – 6.2(2) $\mu_B$ for the Mn spins belonging to the first orbit which sees the moments exclusively in the hexagonal basal plane and between 1.7(2) $\mu_B$ and 5.0(2) $\mu_B$ for the Mn spins belonging to the second orbit where the sinusoidal component along the c-axis is strongest.
The value of the incommensurate magnetic propagation vector $k = [0, 0, 0.4302(1)]$ leads to a spiral with a length of 24.76 Å and a turn angle of 154.87° between spins in neighbouring cells in the c-direction. The $000^{\pm}$ reflection should show up at a 2θ-angle of 4.42° which unfortunately is too close to the primary beam to be seen in our data. No incommensurate is present along the a and b axes modulation and the repeat unit is exactly 8.85 Å.

The resulting structure for MTO is a distorted helical structure with fluctuating values of the Mn moments and directions. Several magnetic Mn sublattices are found to be pure AFM. In spite of the fact that there is only one Mn site in the $Mn_3TeO_6$ crystal structure and all Mn positions in the unit cell have the same crystallographic surrounding, one can not expect in this hexagonal system a perfect AFM structure due to the topography of the Mn-O-Mn bonds. It is important to note that in the case of $Mn_3TeO_6$ there are 3 Mn atoms within a ring (at the same height in the c direction) having equivalent individual bonds giving the classical case of frustration in a triangle.

**4. Conclusions**

$Mn_3TeO_6$ adopts a corundum-related trigonal *R*-centred structure at room temperature, and retains this symmetry down to 5 K. Some key features of the structure are a close packing of strongly distorted hexagonal oxygen layers parallel to (001), with Mn and two distinct Te atoms in the octahedral interstices. Both $TeO_6$ octahedra are fairly regular but the $MnO_6$ octahedron is considerably distorted. Relatively long distances between $Mn^{2+}$ ions are possibly responsible for the low magnetic ordering temperature in this compound.

Below 23 K, $Mn_3TeO_6$ transforms to a magnetically ordered state, however, the ordering is incommensurately modulated. The chain to chain coupling cannot be assigned as pure AFM. The main feature of this magnetic structure is several types of Mn-chains.

Under the action of the incommensurate magnetic propagation vector k = [0, 0, 0.4302(1)] the unique Mn site is split into two magnetically different orbits. Sites, which are crystallographically linked through the inversion centre are no longer coupled from the magnetic point of view. The propagation vector is modulating the spins to a helical spiral with the length of 24.76 Å and a turn angle of 154.87°

**Acknowledgements**

Financial support of this research from the Royal Swedish Academy of Sciences, the Swedish Research Council (VR), the Göran Gustafsson Foundation and the Russian Foundation for Basic Research is gratefully acknowledged. We also gratefully acknowledge the support from N. Sadovskaya and S. Stefanovich during the EDS cation analysis and second harmonic generation testing.

**References**
(1) H. Schmid, J. Phys: Condens. Matter, **20**, 434201 (2008).
(2) N. A. Spaldin, Topics Appl. Phys., **105**, 175 (2007).
(3) *Magnetoelectric Interaction Phenomema in Crystals* Eds. A. J. Freeman and H. Schmid, Gordon and Breach: N.Y., 1975, p. 228.
(4) E. V. Wood and A. E. Austin, Inter. J. Magn., **5**, 303 (1974).
(5) G. A. Smolensky and I. E. Chupis, Sov. Phys.-Uspekhi, **25**, 475 (1982).
(6) D. I. Khomskii, Physics, **2**, 20 (2009).
(7) T. Kimura, T. Goto, H. Shintani, K. Ishizaka, T. Arima, and Y. Tokura, Nature, **426**, 55 (2003).
(8) Y. Tokura, Science, **312**, 1481 (2006).
(9) N. A. Spaldin, S. W. Cheong, and R. Ramesh, Physics Today, **38**, 63 (2010).
(10) R. Ramesh and N. A. Spaldin, Nature Materials, **6**, 21 (2007).
(11) A. Filippetti and N. A. Hill, Phys. Rev. B, **65**, 195120 (2002).
(12) J. Kreisel and M. Kenzelman, Europhys. News, **40**, 17 (2009).
(13) C. N. R. Rao and C. R. Serrao, J. Mater. Chem., **17**, 4931 (2007).
(14) W. Eerenstein, N. D. Mathur, and J. F. Scott, Nature, **442**, 759 (2006).
(15) W. Prellier, M. P. Singh, and P. Murugavel, J. Phys.: Condens. Matter, **117**, R803 (2005).
(16) J. Zupan, D. Kolar, and V. Urbanc, Mater. Res. Bull., **6**, 1353 (1971).
(17) I. Zivkovic, K. Prsa, O. Zaharko, and H. Berger, J. Phys.: Condens. Matter, **22**, 056002 (2010).
(18) K. Y. Choi, P. Lemmens, E. S. Choi, and H. Berger, J. Phys.: Condens. Matter, **20**, 505214 (2008).
(19) G. Bayer, Zeit. Kristallogr., **124**, 131 (1967).
(20) L. I. Kosse, E. D. Politova, and V. V. Chechkin, Izv. AN SSSR, Neorg. Mater. (in Russian), **18**, 1879 (1982).
(21) L. I. Kosse, E. D. Politova, and Yu. N. Venevtsev, Zhurnal Neorg. Khim. (in Russian), **28**, 1689 (1983).
(22) M. Weil, Acta Cryst. E, **62**, i244 (2006).
(23) J. Rodriguez-Carvajal, Physica B, **192**, 55 (1993).
(24) T. B. Zunic and I. Vickovic, J. Appl. Phys., **29**, 305 (1996).
(25) J. Rodriguez-Carvajal, *BASIREPS: a program for calculating irreducible representations of space groups and basis functions for axial and polar vector properties*; 2007.
(26) S. K. Kurtz and T. T. Perry, J. Appl. Phys., **39,** 3798 (l968).
(27) W. Levason, Coord. Chem. Rev., **161**, 33 (1997).
(28) Yu. A. Izyumov, V. E. Naish, and R. P. Ozerov in: *Neutron diffraction of magnetic materials*; P. Consultants Bureau: New York, 1991.

Table 1 Summary of the results of the structural refinements of the Mn$_3$TeO$_6$ sample using XRD and NPD data.

| Experiment | XRD [25] | NPD | NPD |
|---|---|---|---|
| T,K | 295 | 5 | 295 |
| a[Å] | 8.8673(10) | 8.8516(1) | 8.8679(1) |
| c[Å] | 10.6729(12) | 10.6503(2) | 10.6727(2) |
| s.g. | *R*-3 | *R*-3 | *R*-3 |
| **In** | | | |
| *x* | 0.03839(3) | 0.0387(4) | 0.0380(6) |
| *y* | 0.26425(3) | 0.2651(5) | 0.2649(7) |
| *z* | 0.21297(2) | 0.2139(3) | 0.2128(4) |
| *B[Å]$^2$* | 0.59(1) | 0.21(6) | 0.51(7) |
| **Te1** | | | |
| x | 0 | 0 | 0 |
| y | 0 | 0 | 0 |
| z | 0.5 | 0.5 | 0.5 |
| *B[Å]$^2$* | 0.38(1) | 0.46(4) | 0.71(2) |
| **Te2** | | | |
| *x* | 0 | 0 | 0 |
| *y* | 0 | 0 | 0 |
| *z* | 0.5 | 0 | 0 |
| *B[Å]$^2$* | 0.37(1) | 0.08(4) | 0.32(2) |
| **O1** | | | |
| *x* | 0.03069(16) | 0.0297(4) | 0.0300(6) |
| *y* | 0.19625(17) | 0.1967(3) | 0.1961(5) |
| *z* | 0.40283(12) | 0.4028(2) | 0.4030(3) |
| *B[Å]$^2$* | 0.70(2) | 0.16(6) | 0.51(8) |
| **O2** | | | |
| *x* | 0.18277(17) | 0.1832(4) | 0.1833(5) |
| *y* | 0.15620(16) | 0.1573(4) | 0.1561(5) |
| *z* | 0.11053(12) | 0.1111(2) | 0.1113(3) |
| *B[Å]$^2$* | 0.62(2) | 0.20(6) | 0.55(8) |
| $R_p$,% | | 4.65 | 3.16 |
| $R_{wp}$,% | | 6.18 | 4.21 |
| $R_F^2$ / $R_B$(%) | 1.4 | 2.57 | 2.61 |
| $R_{mag}$,% | | 7.68 | |
| $\chi^2$ | | 2.23 | 1.65 |

Table 2. Selected bond lengths [Å] from neutron powder data refinements of the $Mn_3TeO_6$ sample at various temperatures.

| Bonds, Å | | 5K | 295K |
|---|---|---|---|
| Mn | O1 | 2.101(6) | 2.111(6) |
|  | O1 | 2.366(4) | 2.374(6) |
|  | O1 | 2.197(4) | 2.206(6) |
|  | O2 | 2.224(6) | 2.238(8) |
|  | O2 | 2.236(4) | 2.225(5) |
|  | O2 | 2.115(7) | 2.118(8) |
| Te1 | O1 | 1.926(3) | 1.925(4) |
| Te2 | O2 | 1.926(4) | 1.929(5) |

Table 3. Polyhedral analysis of Mn3TeO$_6$ at different temperatures (cn - coordination number, x – shift from centroid, ξ- average bond distance with a standard deviation, V- polyhedral volume, ω- polyhedral volume distortion.

T=295K space group $R\bar{3}$.

| Cation | cn | x(Å) | ξ (Å) | V(Å$^3$) | ω | Valence |
|---|---|---|---|---|---|---|
| Mn | 6 | 0.074 | 2.212+/-0.096 | 12.8(1) | 0.104 | 1.97 |
| Te1 | 6 | 0 | 1.925+/-0.004 | 9.4(1) | 0.007 | 5.89 |
| Te2 | 6 | 0 | 1.929+/-0.005 | 9.5(1) | 0.007 | 5.83 |

T=5K space group $R\bar{3}$

| Cation | cn | x(Å) | ξ (Å) | V(Å$^3$) | ω | Valence |
|---|---|---|---|---|---|---|
| Mn | 6 | 0.077 | 2.206+/-0.096 | 12.7(1) | 0.103 | 1.98 |
| Te1 | 6 | 0 | 1.926+/-0.003 | 9.4(1) | 0.007 | 5.89 |
| Te2 | 6 |  | 1.926+/-0.004 | 9.5(1) | 0.006 | 5.85 |

Table 4. Refined values of the coefficients $C_1$, $C_2$ and $C_3$ for the magnetic phase at 5K (C1 and C2 are not orthogonal to each other but span 120°). Mn1 on x, y, z (0.039, 0.265, 0.213) belongs to orbit 1, Mn2 on –x, -y, -z (0.961, 0.735, 0.787) belongs to orbit 2. Phase between the 2 orbits ϕ = 0.10(1).

| Atom | $C_1$ | $C_2$ | $C_3$ |
|---|---|---|---|
| Mn1 | 5.08(8) | -5.08(8) | 0 |
| Mn2 | 1.37(6) | -1.37(6) | 4.76(9) |

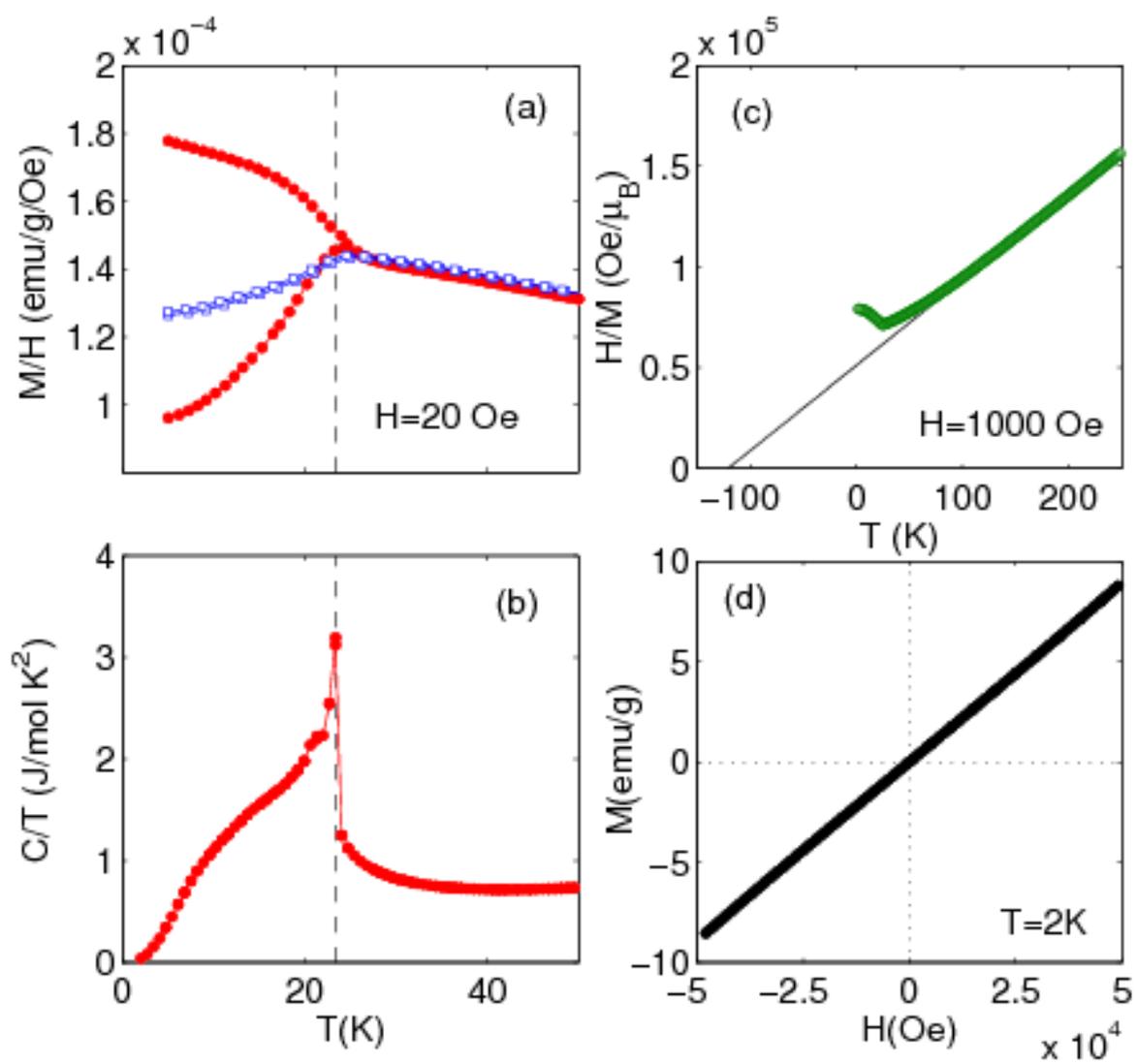

Figure 1

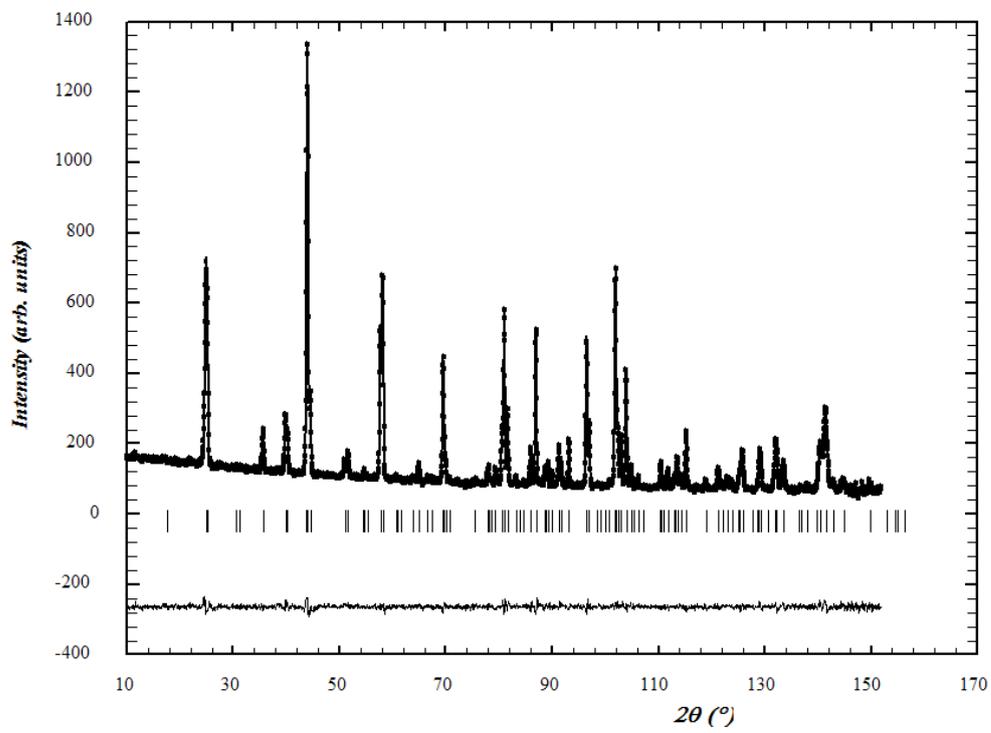

Figure 2

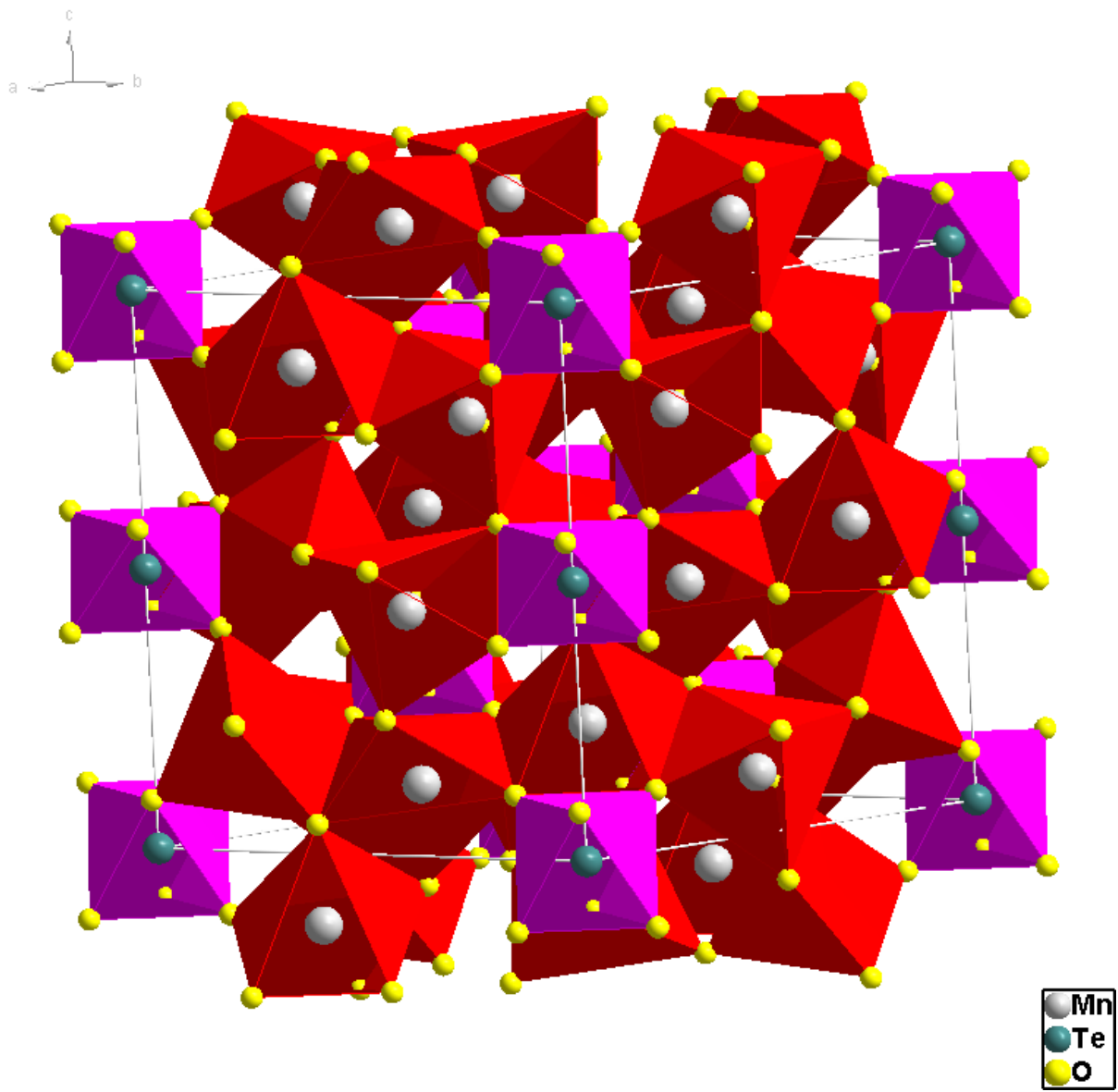

Figure 3

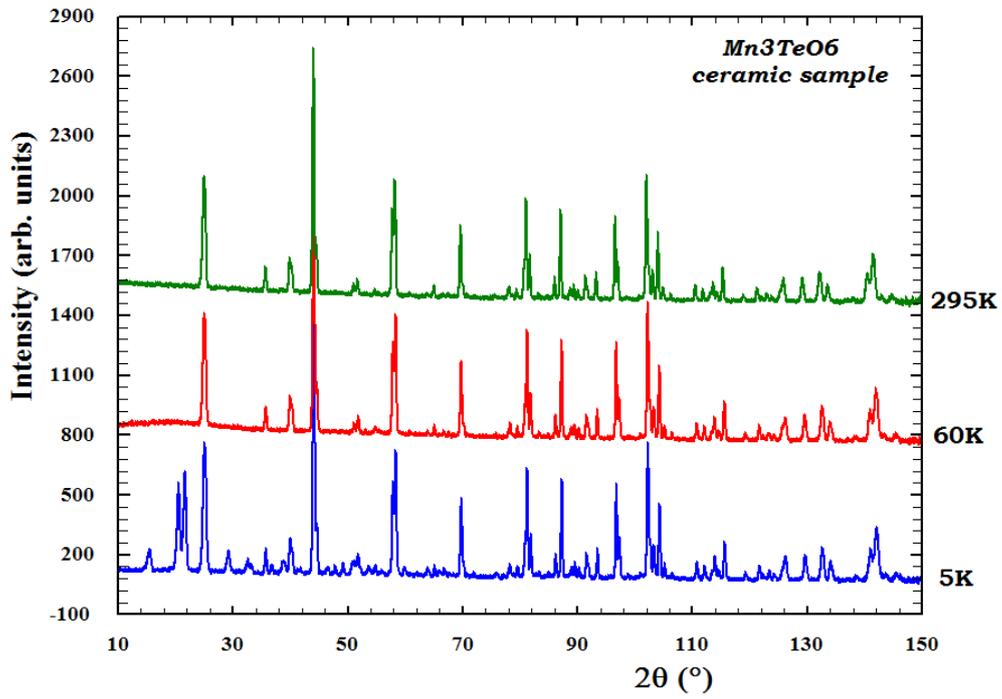

Figure 4

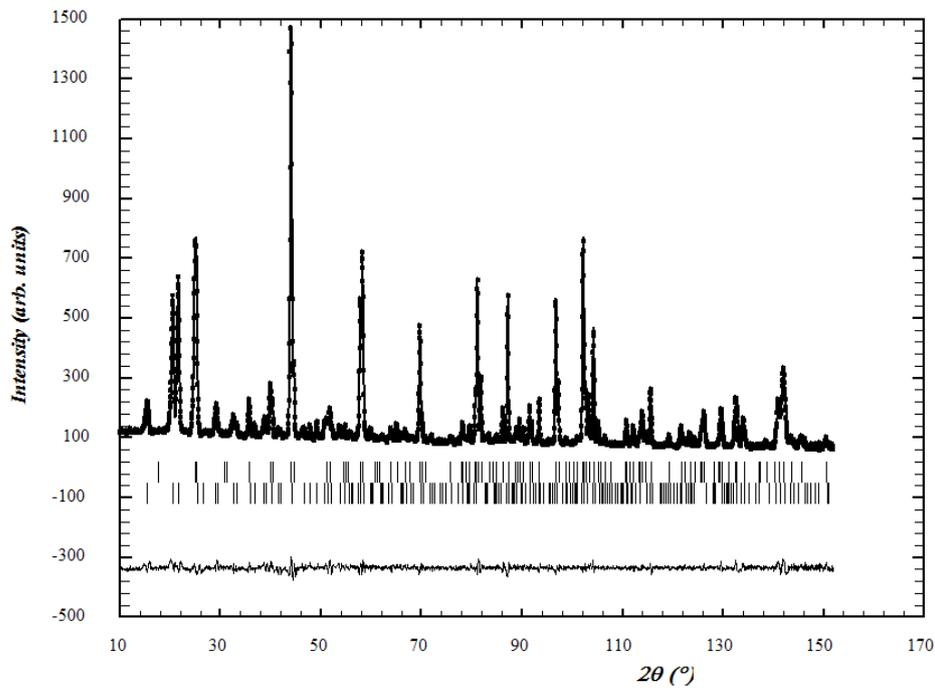

Figure 5

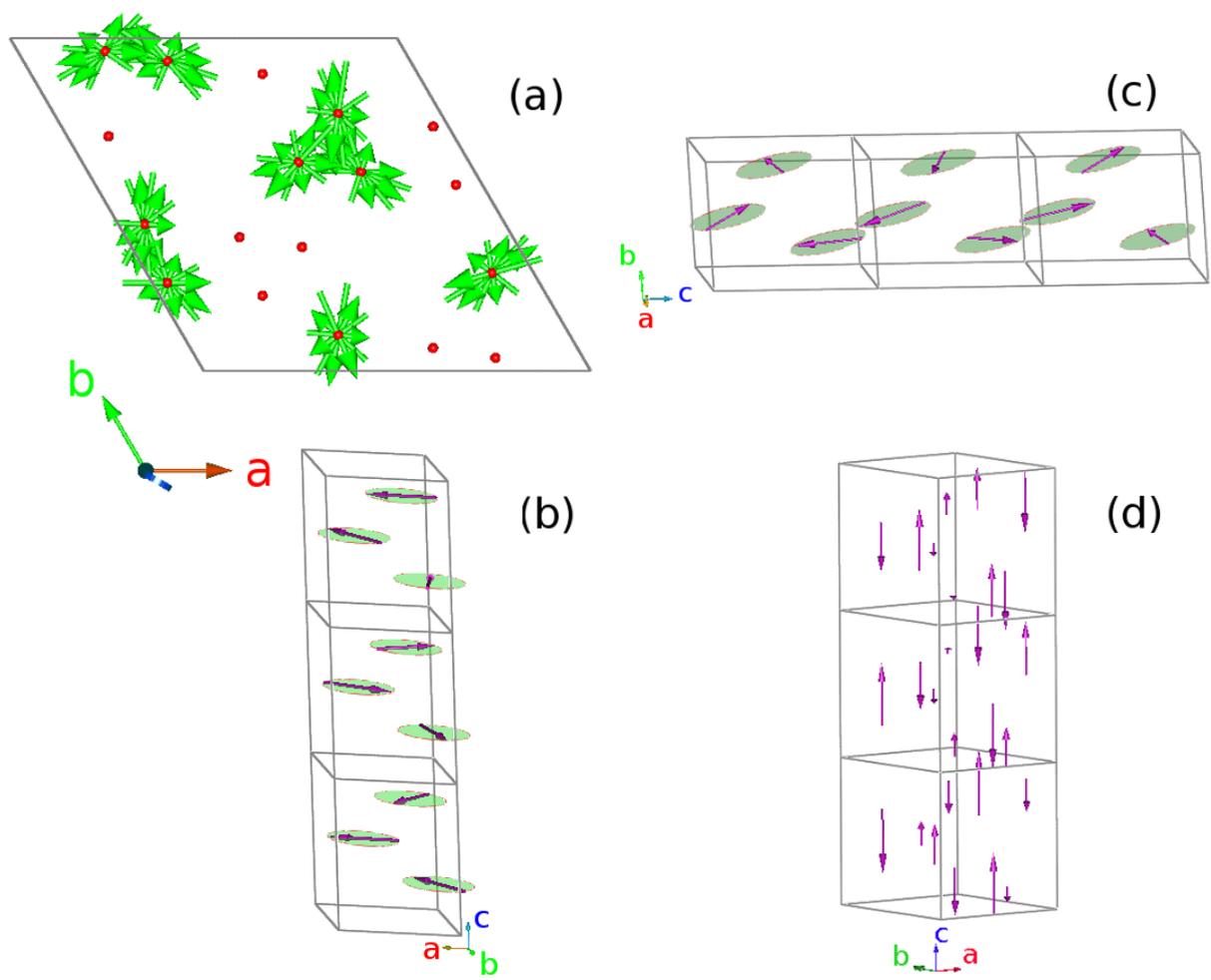

Figure 6

**Figure captions**

**Figure 1** (Color online) Temperature dependence of the (a) ZFC/FC susceptibility M/H and (b) heat capacity C (plotted as C/T) of $Mn_3TeO_6$ single-crystal (filled symbols). The corresponding ZFC/FC data for the polycrystalline sample used in NPD is added in (a) using open symbols. The vertical dash line marks $T_N$ near 23 K. Panel (c) shows the Curie-Weiss fit of the high-field (H=1000 Oe) single-crystal data, while panel (d) shows an hysteresis curve at T = 2K recorded for the single-crystal.

**Figure 2** The observed, calculated, and difference plots for the fit to the NPD pattern of $Mn_3TeO_6$ after Rietveld refinement of the nuclear structure at 295K.

**Figure 3** (Color online) Polyhedral representation of crystal structure of $Mn_3TeO_6$.

**Figure 4** (Color online) Temperature evolution of NPD patterns of $Mn_3TeO_6$.

**Figure 5** The observed, calculated, and difference plots for the fit to the NPD patterns of $Mn_3TeO_6$ after Rietveld refinement of the nuclear and magnetic structure at 5K.

**Figure 6** (Color online) Magnetic structure of $Mn_3TeO_6$ at 5K: (a) the elliptical envelope of the Mn(1) orbit perpendicular to the propagation vector (b) the variation of the magnetic moments in the Mn(1) sites along the helical spiral (c) the variation of the magnetic moments in the Mn(2) sites along the propagation vector (d) the sine wave character of the Mn(2) orbit along the c-axis.